\documentclass[aps,prl,twocolumn,superscriptaddress,nofootinbib,longbibliography]{revtex4-2}

\usepackage{amsmath,amssymb,amsfonts,mathtools,bm}
\usepackage{graphicx}
\usepackage{hyperref}
\usepackage{xcolor}
\usepackage{physics}
\usepackage{bbm}
\usepackage{booktabs}

\begin{document}

\title{Exact Holographic Kinematics in AdS/CFT}

\author{Haitang Yang}
\email{hyanga@scu.edu.cn}
\affiliation{College of Physics, Sichuan University, Chengdu, 610065, China}

\begin{abstract}
We propose that holography contains an exact kinematic sector distinct from holographic dynamics.
The appropriate setting for this sector is a CFT on an open solid torus in the Weyl frame.
The open solid torus introduces an intrinsic scale, and the Weyl frame makes this scale manifest as an extra bulk direction.
The resulting bulk-boundary pairs are exact and finite: no cutoff, large-$N$ limit, strong-coupling assumption, or heavy-operator approximation is required.
The AdS geometry appearing in this sector should be understood as a kinematic geometry;
only in special CFTs and appropriate limits is it promoted to a dynamical semiclassical  gravitational bulk theory.
The standard boundary-anchored dictionary entries are recovered  as singular limits.
As a   demonstration, we show that  Weyl-frame two-point functions provide
a replica-free definition of entanglement entropy.
\end{abstract}

\maketitle

\paragraph{Introduction and discussion.---}
A central lesson of holography is that bulk geometry is encoded in boundary data \cite{Maldacena:1997re,Gubser:1998bc,Witten:1998qj}.
The extra bulk direction is often associated with the renormalization group scale of the boundary theory.
In practice, however, the standard formulation is usually organized through asymptotic relations: bulk quantities are brought to the conformal boundary, divergent factors are removed by cutoff-dependent prescriptions, and the resulting finite parts are matched with boundary observables.
Semi-classical limits such as large-$N$, strong-coupling or heavy-operator approximation are usually imposed during this process.
This asymptotic organization becomes especially explicit in boundary-anchored geodesic relations and in the Ryu--Takayanagi formula \cite{Ryu:2006bv,Hubeny:2007xt}.

This raises a conceptual question.
Many familiar holographic pairs contain a universal kinematic core.
On the CFT side, the coordinate dependence of two- and three-point functions,
the kinematic factors of higher-point functions,
and the geometric dependence of entanglement entropies
are completely fixed by conformal symmetry.
Those quantities are independent of the details of specific CFTs,
up to theory-dependent data such as scaling dimensions, OPE coefficients, central charges, and vacuum energies.
On the AdS side, the corresponding objects are geometric invariants such as geodesic lengths, hyperbolic volumes, and minimal surfaces.
\emph{Logically, if the kinematic sector matches in the dynamical limits, since they are independent of dynamics, 
their matching should be independent of the dynamical limits. This suggests that a part of holography should be understood as kinematics rather than dynamics.} Since kinematics is fixed by symmetry, there is no intrinsic reason for this sector to require large $N$, strong coupling, or heavy-operator limits, which belong instead to the dynamical or semiclassical regime.

Moreover, in a UV-complete theory such as a CFT, cutoffs should not be regarded as intrinsic ingredients of the theory; they should arise only from singular limiting procedures.

\begin{figure}[h]
\centering
\includegraphics[scale=0.35]{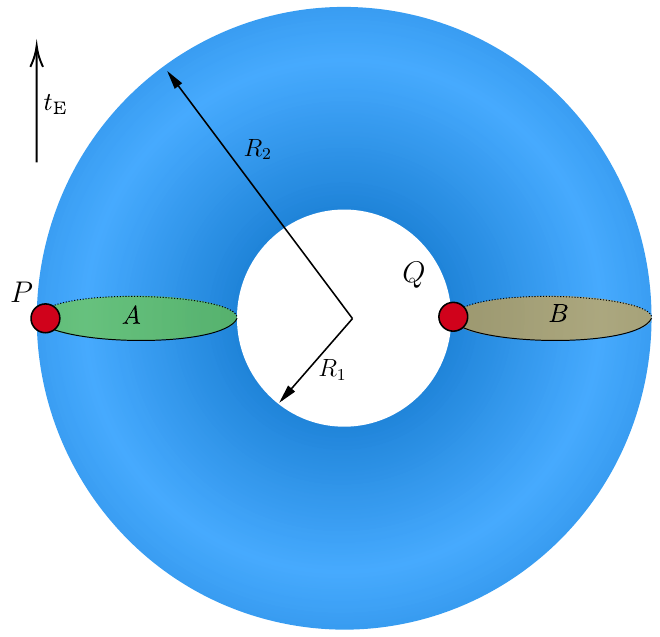}
\caption{The CFT$_D$ lives on the flat open solid torus $S^1\times\mathbb B^{D-1}$ with Euclidean flat metric.
The Euclidean time direction is upward.
This geometry prepares a pure state with disjoint subregions $A$ and $B$.
As shown in Ref.~\cite{Jiang:2025jnk}, the disjoint entanglement entropy $S_{\rm disj}(A:B)$ is finite and exactly matches the bulk EWCS $E_W(A:B)$.
\label{fig:solid-torus}}
\end{figure}

\begin{figure}[h]
\centering
\includegraphics[scale=0.3]{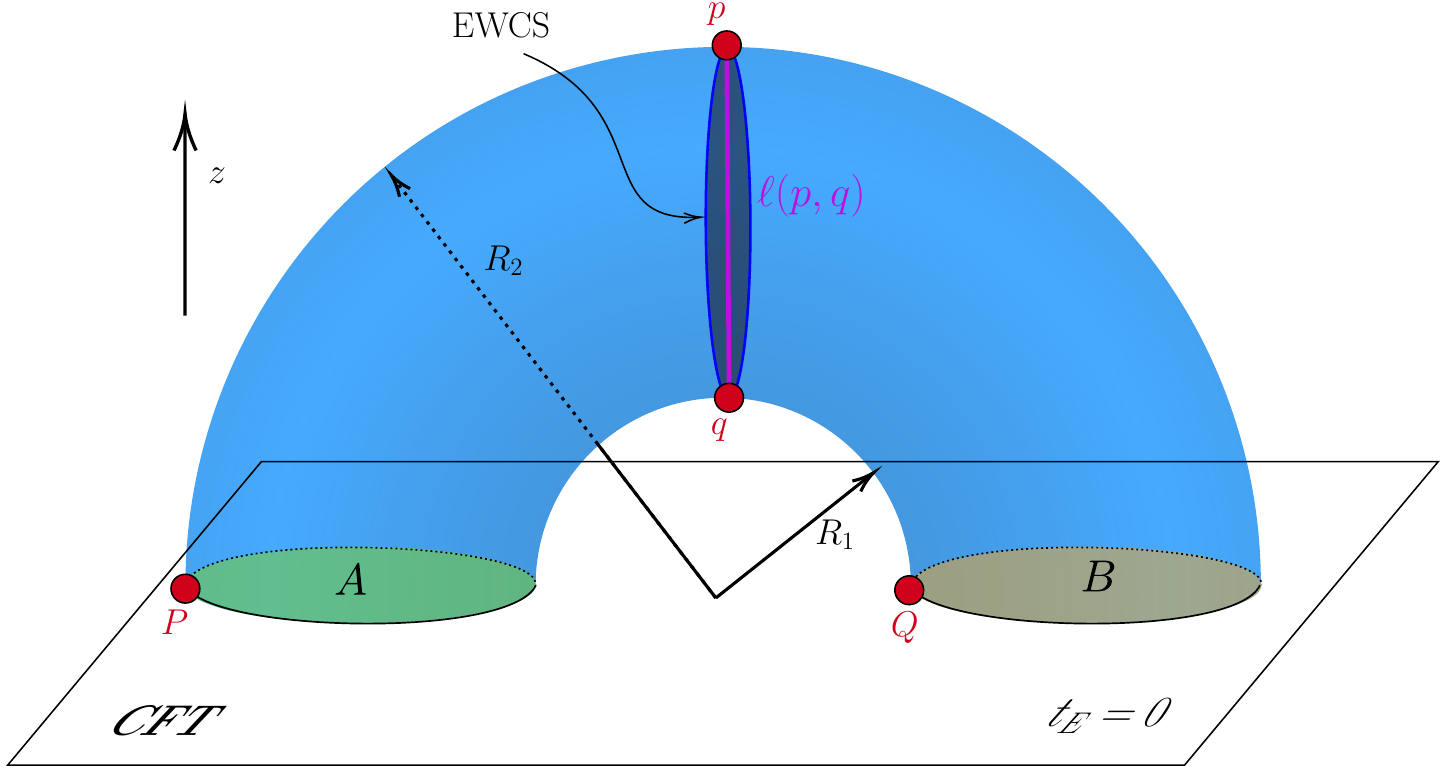}
\caption{Euclidean AdS$_{D+1}$ geometry $\mathrm d s^2_{\rm EAdS}= (\mathrm d t_E^2+\mathrm d z^2+\sum_I\mathrm d x_I^2)/z^2$, with the bulk direction $z$ upward.
The points $P$ and $Q$ are the correlator points in CFT$_D$.
The shaded sphere denotes the EWCS of $A$ and $B$.
The points $p$ and $q$ are antipodal points of the EWCS.
\label{fig:RT}}
\end{figure}

In previous works \cite{Jiang:2025jnk,Jiang:2026juf}, starting from a CFT on an open solid torus, we established two exact bulk-boundary pairs:
\begin{equation}
S_{\rm disj}(A:B) \equiv E_W(A:B),
\label{eq:exact-pairs-I}
\end{equation}
\begin{equation}
G_{\mathcal W}(P,Q) \equiv C_\Delta \Big[2\cosh\frac{\ell(p,q)}{2}\Big]^{-2\Delta}.
\label{eq:exact-pairs-II}
\end{equation}
Here $S_{\rm disj}(A:B)$ is the entanglement entropy between disjoint complementary regions $A$ and $B$ on an open solid torus, and $E_W(A:B)$ is the corresponding entanglement wedge cross-section (EWCS) in AdS.
The quantity $G_{\mathcal W}(P,Q)$ is the Weyl-frame two-point function between $P$ and $Q$ on the open solid torus, while $\ell(p,q)$ is the length of a finite geodesic lying entirely in the bulk, with the bulk points $p$ and $q$ determined by the boundary solid-torus geometry.
Figures~\ref{fig:solid-torus} and \ref{fig:RT} illustrate these quantities.
Both pairs are finite, exact and require no cutoff, large-$N$ limit, strong-coupling assumption, or heavy-operator approximation.
The usual boundary-anchored relations are recovered only by taking singular limits.

How could the exact pairs be possible?
The critical mechanism is the open solid torus in the Weyl frame:
{\it the solid torus introduces an intrinsic scale, and the Weyl frame makes this scale manifest}.
This provides a clean way in which a boundary scale is promoted to an extra bulk direction.
The resulting observables are finite and well defined from the outset, while the usual divergent expressions arise only after degeneration.

These observations motivate the following distinction:
\begin{itemize}
\item \emph{Holographic kinematics}: universal structures fixed by conformal covariance.
\item \emph{Holographic dynamics}: model-dependent structures determined by the spectrum, OPE data, interactions, and semiclassical gravitational dynamics.
\end{itemize}
Large-$N$ factorization, strong coupling, and heavy-operator limits are not the origin of holographic kinematics; they are needed only when kinematic geometry is promoted to classical gravitational dynamics.
This separation is summarized in Table~\ref{tab:kinematics-dynamics}.

\begin{table*}[t]
\centering
\renewcommand{\arraystretch}{1.2}
\begin{tabular*}{\textwidth}{p{0.18\textwidth} p{0.36\textwidth} p{0.36\textwidth}}
\toprule
\textbf{Layer} & \textbf{CFT} & \textbf{AdS} \\
\midrule
Kinematics
&
Conformal/Weyl geometry; fixed correlator dependence; cross-ratios.
&
Geodesic lengths; hyperbolic volumes; minimal surfaces.
\\
Theory data
&
$\Delta_i$, $C_{ijk}$, central charges, $E_{\rm vac}$, spectrum.
&
Masses, couplings, $G_N$, normalizations.
\\
Dynamics
&
Model-dependent OPE data, partition functions, thermal spectra.
&
Equations of motion, interactions, backreaction, black holes.
\\
\bottomrule
\end{tabular*}
\caption{
Holographic kinematics should be distinguished from holographic dynamics.
Exact pairs relate CFT conformal/Weyl kinematics to finite AdS geometric invariants.
}
\label{tab:kinematics-dynamics}
\end{table*}
Our goal in this work is not to address the dynamical sector.
Rather, we isolate the part of holography controlled by symmetry.
We claim that holographic duality contains an exact kinematic sector independent of the usual dynamical limits.
This distinction is essential: kinematics is fixed by symmetry, whereas dynamics is only constrained by it.
Conventional derivations often mix the two because geometry is usually extracted in a semiclassical regime through large $N$, strong coupling, or heavy-operator limits.
By contrast, once scale is geometrized in the appropriate way---in particular, through the open solid torus in the Weyl frame---the kinematic dictionary can be exact without invoking any such dynamical approximation.

\paragraph{Kinematic AdS geometry versus dynamical bulk.}
The distinction between holographic kinematics and holographic dynamics is important for
interpreting the above exact pairs. We do not claim that every CFT with the same conformal
kinematics possesses a semiclassical Einstein bulk. Rather, conformal symmetry and Weyl
covariance naturally organize CFT observables into invariants that admit an AdS-geometric
representation. This representation is kinematic: it is available for any CFT in the same way
that two- and three-point functions are fixed by conformal symmetry up to theory data such
as $\Delta_i$, $C_{ijk}$, central charges, and vacuum energies. The additional assumptions
usually invoked in holography, such as large $N$, sparse spectrum, strong coupling, or
heavy-operator limits, are not needed to define this kinematic geometry. They are needed
only when one further promotes the kinematic AdS geometry to a dynamical semiclassical
bulk spacetime.

This viewpoint also clarifies the role of the exact pairs. For a generic CFT, the geodesics,
hyperbolic volumes, and minimal surfaces appearing in the solid-torus Weyl frame should be
understood as objects in an exact kinematic AdS geometry. For a holographic CFT, these same
objects can coincide with the corresponding structures in the physical bulk. Thus the exact
pair program does not assert that conformal kinematics alone determines the full bulk
dynamics. Instead, it identifies the universal geometric sector that precedes, and is logically
distinct from, the dynamical gravitational realization.

As a significant demonstration of the power of separating kinematics from dynamics,
we show that,
for the spherical configurations considered here, the entanglement entropy can be defined from Weyl-frame two-point data through the chain
\begin{widetext}
\begin{equation}
G_{\mathcal W}(P,Q)
\overset{\rm Pair\ II}{\longrightarrow}
\ell(p,q)
\longrightarrow
g_{\mu\nu}
\longrightarrow
E_W
\overset{\rm Pair\ I}{\longrightarrow}
S_{\rm disj}
\longrightarrow
S_{\rm adj},
\end{equation}
\end{widetext}
where $S_{\rm adj}$ is the usual divergent entanglement entropy between adjacent regions.
Every step is exact and finite until the final adjacent singular limit.
No large-$N$, strong-coupling, or heavy-operator limit is taken.
This provides a replica-free route to the CFT entanglement entropy.

\paragraph*{Setup: CFT$_D$ on an open solid torus.}
Our setting is Euclidean throughout.
For a CFT in $D$-dimensional spacetime with flat metric
$ \mathrm d s_E^2=\mathrm d t_E^2+\mathrm d y^2+
\sum_{I=1}^{D-2}\mathrm d x_I^2, $
we consider quantum fields living on an open solid torus $S^1\times\mathbb B^{D-1}$,
\begin{equation}
\mathcal B_D=\left\{\left(\sqrt{t_E^2+y^2}-\frac{R_2+R_1}{2}\right)^2+
\sum_{I=1}^{D-2}x_I^2<\left(\frac{R_2-R_1}{2}\right)^2\right\},
\end{equation}
with $R_2>R_1>0$, as shown in Fig.~\ref{fig:solid-torus}.
Here $\mathbb B^{D-1}$ denotes a $(D-1)$-dimensional open ball.
Since the solid torus is open, it carries the induced flat metric.

\paragraph{Weyl-frame correlator $\to$ bulk geodesic.---}
Referring to Figs.~\ref{fig:solid-torus} and \ref{fig:RT}, let
$P(r=R_2,\theta_P=\pi,x_I=0)$ and $Q(r=R_1,\theta_Q=0,x_I=0)$ be two points in the solid-torus CFT.
Let $G_{\mathcal W}(P,Q)$ denote the Weyl-frame two-point function of a scalar primary operator $\mathcal O$ with dimension $\Delta$.
In Ref.~\cite{Jiang:2026juf}, we showed the exact pairing \eqref{eq:exact-pairs-II}.

Although the pairing is illustrated for the symmetric solid torus,
it extends to generic asymmetric sphere configurations by expressing the geodesic length in a conformally invariant form,
\begin{equation}
\ell(p,q)=\log\frac{\sqrt{\varrho+1}+\sqrt2}{\sqrt{\varrho+1}-\sqrt2},
\label{eq:GL-rho}
\end{equation}
where the inversive product $\varrho$ is the conformal invariant associated with two spheres of centers $\vec x,\vec x'$ and radii $r,r'$,
\begin{equation}
\varrho=\left|\frac{r^2+r^{\prime 2}-|\vec x-\vec x'|^2}{2rr'}\right|.
\label{eq:inversive-product}
\end{equation}
Eq.~\eqref{eq:exact-pairs-II} can be inverted to obtain
\begin{equation}
\ell(p,q)=
2\operatorname{arccosh}\!\left[
\frac{1}{2}
\left(\frac{C_\Delta}{G_{\mathcal W}(P,Q)}\right)^{\frac{1}{2\Delta}}
\right].
\label{eq:ellfromGW}
\end{equation}
This relation is exact.
It does not rely on a geodesic approximation or any semiclassical dynamical limit.

\paragraph{Geodesics $\to$ metric.---}
From Fig.~\ref{fig:RT}, the $(D+1)$-dimensional coordinates of the bulk points $p$ and $q$,
are fixed as the antipodal points of the EWCS by the solid-torus geometry.
The same statement holds for generic asymmetric sphere configurations after conformal transformations.
Denote the bulk coordinates by $Y^\mu=(z,t_E,\vec x)$ and $Y^{\prime\mu}=(z',t_E',\vec x')$.
In terms of Synge's world function $\sigma(Y,Y')=\frac12\ell^2(Y,Y')$, one has \cite{Synge:1960ueh}
\begin{equation}
g_{\mu\nu}(Y)=
-\lim_{Y'\to Y}\partial_\mu\partial_{\nu'}\sigma(Y,Y')
=-\lim_{Y'\to Y}\partial_\mu\partial_{\nu'}\left(\frac12\ell^2\right).
\label{eq:GtoM1}
\end{equation}
The derivatives act on different endpoints of the geodesic.

\paragraph{Metric $\to$ minimal surfaces.---}
The minimal surface relevant here is the EWCS.
For a generic shape, computing the minimal surface from the metric might be difficult.
For sphere configurations, however, it is sufficient to work in a symmetric representative and then use conformal transformations to obtain the corresponding asymmetric cases.
Referring to Fig.~\ref{fig:RT}, the codimension-two EWCS has volume \cite{Jiang:2025jnk}
\begin{eqnarray}
\operatorname{Vol}(E_W) &=&
\pi^{\frac{D-2}{2}}
\frac{\Gamma\left(\frac{D}{2}\right)}{\Gamma(D)}
\left(\frac{2\sqrt2}{\sqrt{\varrho+1}-\sqrt2}\right)^{D-1}\nonumber\\
&\times&
{}_{2}F_{1}\!\left(D-1,\frac{D}{2};D;-
\frac{2\sqrt2}{\sqrt{\varrho+1}-\sqrt2}\right).
\label{eq:Gen-ee}
\end{eqnarray}
This expression has already been written in terms of the conformal invariant $\varrho$, and is therefore applicable to asymmetric sphere configurations.
Here ${}_{2}F_1$ is the standard hypergeometric function.

\paragraph{Minimal surface $\to$ entanglement entropy.---}
In Ref.~\cite{Jiang:2025jnk}, we showed that the disjoint entanglement entropy between $A$ and $B$ equals the EWCS volume with a theory-dependent normalization,
\begin{equation}
S_{\rm disj}(A:B)=4\pi|\mathcal E_{\rm vac}|\operatorname{Vol}(E_W).
\label{eq:S-V}
\end{equation}
In a semiclassical holographic realization, the normalization is conventionally identified with
\(4\pi|\mathcal E_{\rm vac}|=1/(4G_{D+1})\). In the present kinematic formulation, however, \(\mathcal E_{\rm vac}\) should be viewed as CFT theory data.

\paragraph{Correlator $\to$ entanglement entropy.---}
Combining Eqs.~\eqref{eq:GL-rho}, \eqref{eq:ellfromGW}, \eqref{eq:Gen-ee}, and \eqref{eq:S-V}, we obtain an exact relation between the Weyl-frame solid-torus correlator and the disjoint entanglement entropy defined by the same solid torus:
\begin{widetext}
\begin{equation}
\boxed{
S_{\rm disj}(A:B) = 4|\mathcal E_{\rm vac}|\,
\pi^{\frac{D}{2}}\frac{\Gamma\left(\frac{D}{2}\right)}{\Gamma(D)}
\left(e^{2\operatorname{arccosh}\!\left[
\frac{1}{2}
\left(
\frac{C_\Delta}{G_{\mathcal W}(P,Q)}
\right)^{\frac{1}{2\Delta}}
\right]}-1\right)^{D-1}
{}_{2}F_{1}\!\left(D-1,\frac{D}{2};D;
1-e^{2\operatorname{arccosh}\!\left[
\frac{1}{2}
\left(
\frac{C_\Delta}{G_{\mathcal W}(P,Q)}
\right)^{\frac{1}{2\Delta}}
\right]}
\right).
}
\label{eq:EE-2pt}
\end{equation}
\end{widetext}

Eq.~\eqref{eq:EE-2pt} provides a replica-free definition of the disjoint entanglement entropy for the spherical configurations considered here.
It is a direct and finite relation from Weyl-frame two-point data to entanglement entropy.
Unlike the usual semiclassical route, this map is not derived by assuming that a saddle-point bulk geometry already exists.
Rather, it follows from the exact kinematic dictionary supplied by the open solid torus and the Weyl frame.

\begin{figure}[h]
\centering
\includegraphics[scale=0.35]{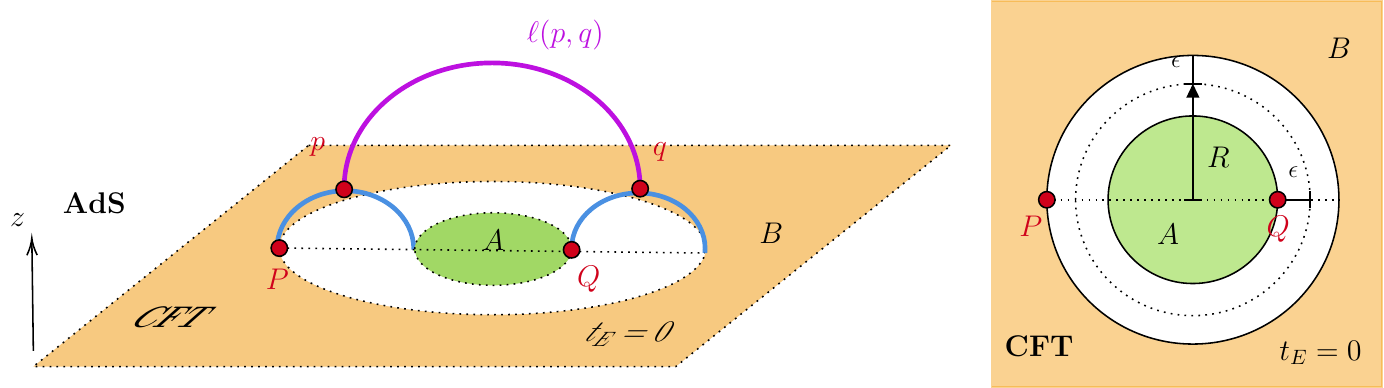}
\caption{The cavity configuration.
The $(D-1)$-ball $A$ of radius $R-\epsilon$ lies inside a spherical cavity of radius $R+2\epsilon$, while the region $B$ lies outside the cavity.
Here $\epsilon$ plays the role of the UV cutoff in the adjacent limit.
The dual bulk geodesic is the purple curve $\ell(p,q)$.
\label{fig:limit}}
\end{figure}

\paragraph{Boundary limit and recovery of the standard formula.---}
The conventional boundary expression is recovered as a singular limit.
The cavity configuration in Fig.~\ref{fig:limit} makes it transparent how the usual divergent entropy $S_{\rm adj}$ between adjacent regions arises from the singular limit $\epsilon\to0$.
In this configuration, $|x_P-x_Q|=2R+2\epsilon$.
Eq.~\eqref{eq:inversive-product} gives
$\varrho=(R^2+\epsilon^2)/(R^2-\epsilon^2)$.
Using Eqs.~\eqref{eq:GL-rho}, \eqref{eq:ellfromGW} and \eqref{eq:EE-2pt}, we find
\begin{widetext}
\begin{equation}
S_{\rm adj}^{(D)}=\lim_{\epsilon\to0}S_{\rm disj}(A:B)
=\lim_{\epsilon\to0}
4\pi^{\frac{D}{2}}|\mathcal E_{\rm vac}|\,
\frac{\Gamma\left(\frac{D}{2}\right)}{\Gamma(D)}
\left(\frac{2\sqrt{R^2-\epsilon^2}}{R-\sqrt{R^2-\epsilon^2}}\right)^{D-1}
{}_{2}F_{1}\!\left(D-1,\frac{D}{2};D;
-\frac{2\sqrt{R^2-\epsilon^2}}{R-\sqrt{R^2-\epsilon^2}}
\right).
\label{eq:adj-EE-2pt}
\end{equation}
\end{widetext}
For $D=2$ and $\mathcal E_{\rm vac}^{(2)}=-c/(24\pi)$, this gives
$S_{\rm adj}^{(2)}=(c/3)\log(|x_P-x_Q|/\epsilon)+\mathcal O(\epsilon^0)$, in agreement with the standard CFT$_2$ result \cite{Holzhey:1994we,Calabrese:2004eu,Calabrese:2009qy}.
For higher dimensions, the same limit reproduces the area law \cite{Jiang:2025jnk}.

This limit is conceptually important.
It shows that the familiar boundary formula is not the fundamental starting point, but rather the singular shadow of a finite relation.
The natural order is therefore reversed: one should begin with finite observables on the open solid torus and obtain the conventional boundary expressions only by degeneration.

In addition to spherical entangling surfaces, the construction is exact for reflection-symmetric non-spherical solid-torus configurations, where the corresponding volume can be computed in the Weyl-frame geometry.
For arbitrary shapes, the same kinematic prescription defines the corresponding geometric entropy functional.
If the RT/EWCS identification is assumed, the construction extends to arbitrary entangling regions.

\paragraph{Interpretation.---}
Our result suggests a revised conceptual picture of holographic kinematics.
The conventional adjacent or asymptotic observables are not the most primitive objects.
They are singular descendants of finite quantities naturally defined on the open solid torus.
From this viewpoint, the open solid torus is not a special subsector of the usual setup.
Rather, it is the finite parent formulation from which the standard expressions emerge by degeneration.

This interpretation parallels the relation between finite disjoint configurations and their adjacent limits: the finite object is primary, while the adjacent expression arises only after degeneration.
The same logic applies to two-point functions and to entanglement entropy.
All these results suggest:

\begin{widetext}
\center
\boxed{\hbox{Do not regularize the singular object; find its finite parent.}}
\end{widetext}

\paragraph{Outlook: bootstrap.}
The same kinematic viewpoint may also be useful for the conformal bootstrap.
For higher-point correlators, conformal symmetry fixes the kinematic dependence up to functions of cross-ratios, while the spectrum and OPE coefficients encode the dynamical data.
In the solid-torus Weyl frame, these cross-ratios can be written in terms of finite geodesic invariants.
Crossing symmetry may therefore be interpreted as the equivalence of different OPE decompositions of the same underlying bulk-kinematic configuration.
This suggests a possible geometric formulation of the bootstrap, where exact holographic kinematics provides the universal stage and the bootstrap constrains the allowed dynamics.
We leave this direction for future work.

\vspace*{3.0ex}
\begin{acknowledgments}
\paragraph*{Acknowledgments.}
We are indebted to Bo Feng, Xin Gao and Xin Jiang for very useful comments which helped us to improve the manuscript substantially.
This work is supported in part by NSFC (Grant No. 12275184).
\end{acknowledgments}

\bibliographystyle{unsrturl}
\bibliography{ref202605}

\end{document}